\documentstyle[12pt]{article}
\begin{document}

\author{C. Bizdadea\thanks{%
e-mail address: bizdadea@central.ucv.ro} \\
Department of Physics, University of Craiova\\
13 A. I. Cuza Str., Craiova RO-1100, Romania}
\title{Consistent interactions in the Hamiltonian BRST formalism }
\maketitle

\begin{abstract}
A Hamiltonian BRST deformation procedure for obtaining consistent
interactions among fields with gauge freedom is proposed. The general theory
is exemplified on the three-dimensional Chern-Simons model.

\end{abstract}

\section{Introduction}

The analysis of consistent interactions that can be introduced among fields
with gauge freedom without changing the number of gauge symmetries \cite{1}--%
\cite{4} has been transposed lately at the level of the deformation of the
master equation \cite{5} from the antifield-BRST formalism \cite{6}--\cite
{10}. This cohomological deformation technique has been applied, among
others, to Chern-Simons models \cite{5}, Yang-Mills theories \cite{11} and
two-form gauge fields \cite{12}--\cite{12a}. In this light, the
antifield-BRST method was proved to be an elegant tool for investigating the
problem of consistent interactions.

Recently, a Hamiltonian analysis of anomalies has been given \cite{12b}.
Moreover, in a very interesting paper \cite{12c}, there has been established
the precise relation of the local BRST cohomologies in both Lagrangian and
Hamiltonian formalisms (see Theorem 6 from this reference). The procedures
developed within these papers strongly stimulate a Hamiltonian BRST approach
to other interesting problems. 

In this letter we analyze the problem of constructing consistent
interactions among fields with gauge freedom in the framework of the
Hamiltonian BRST formalism \cite{10}, \cite{13}--\cite{17}. Our strategy
includes two main steps: (i) initially, we show that the problem of
introducing consistent interactions among fields with gauge freedom can be
reformulated as a problem of deforming the BRST charge and the
BRST-invariant Hamiltonian with respect to a given ``free'' theory, and
consequently we deduce the general equations that govern these two types of
deformations; (ii) next, on behalf of the relationship between the
Hamiltonian and antifield BRST formalisms for constrained systems we prove
that the general equations possess solution. In the sequel, we reformulate
the general equations in a manner that accounts for locality, and
subsequently illustrate our general procedure in the case of
three-dimensional Chern-Simons models. Finally, we remark that our method
combined with the results in \cite{12c} may simplify the computation of
local Lagrangian BRST cohomologies in some cases of interest.

\section{General equations of the Hamiltonian deformation approach}

We begin with a system described by the canonical variables $z^{A}$, subject
to the first-class constraints 
\begin{equation}
G_{a_{0}}\left( z^{A}\right) \approx 0,\;a_{0}=1,\ldots ,M_{0},  \label{1}
\end{equation}
which are assumed to be $L$-stage reducible 
\begin{equation}
G_{a_{0}}Z_{\;\;a_{1}}^{a_{0}}=0,\;a_{1}=1,\ldots ,M_{1},  \label{2}
\end{equation}
\begin{equation}
Z_{\;\;a_{k-1}}^{a_{k-2}}Z_{\;\;a_{k}}^{a_{k-1}}\approx 0,\;a_{k}=1,\ldots
,M_{k,\;}k=2,\ldots ,L,  \label{3}
\end{equation}
and suppose that there are no second-class constraints in the theory. The
Grassmann parities of the canonical variables and first-class constraints
are respectively denoted by $\varepsilon \left( z^{A}\right) =\varepsilon
_{A}$ and $\varepsilon \left( G_{a_{0}}\right) =\varepsilon _{a_{0}}$. We
denote the first-class Hamiltonian by $H_{0}$, such that the gauge algebra
is expressed by 
\begin{equation}
\left[ G_{a_{0}},G_{b_{0}}\right]
=G_{c_{0}}C_{\;\;a_{0}b_{0}}^{c_{0}},\;\left[ H_{0},G_{a_{0}}\right]
=G_{b_{0}}V_{\;\;a_{0}}^{b_{0}}.  \label{4}
\end{equation}
It is known that a constrained Hamiltonian system can be described by the
action 
\begin{equation}
S_{0}\left[ z^{A},u^{a_{0}}\right] =\int\limits_{t_{1}}^{t_{2}}dt\left(
a_{A}\left( z\right) \dot{z}^{A}-H_{0}-G_{a_{0}}u^{a_{0}}\right) ,  \label{5}
\end{equation}
where the Grassmann parities of the Lagrange multipliers are given by $%
\varepsilon \left( u^{a_{0}}\right) =\varepsilon _{a_{0}}$. In (\ref{5}), $%
a_{A}\left( z\right) $ is the one-form potential that gives the symplectic
two-form $\omega _{AB}=\left( -\right) ^{\varepsilon _{A}+1}\frac{\partial
^{L}a_{A}}{\partial z^{B}}-\left( -\right) ^{\varepsilon _{B}\left(
\varepsilon _{A}+1\right) }\frac{\partial ^{L}a_{B}}{\partial z^{A}}$, whose
inverse, $\omega ^{AB}$, corresponds to the fundamental Dirac brackets $%
\left[ z^{A},z^{B}\right] =\omega ^{AB}$. Action (\ref{5}) is invariant
under the gauge transformations 
\begin{equation}
\delta _{\epsilon }z^{A}=\left[ z^{A},G_{a_{0}}\right] \epsilon
^{a_{0}},\;\delta _{\epsilon }u^{a_{0}}=\dot{\epsilon}^{a_{0}}-V_{\;%
\;b_{0}}^{a_{0}}\epsilon ^{b_{0}}-C_{\;\;b_{0}c_{0}}^{a_{0}}\epsilon
^{c_{0}}u^{b_{0}}-Z_{\;\;a_{1}}^{a_{0}}\epsilon ^{a_{1}}.  \label{6}
\end{equation}

In order to generate consistent interactions at the Hamiltonian level, we
deform the action (\ref{5}) by adding to it some interaction terms 
\begin{equation}
S_{0}\rightarrow \tilde{S}_{0}=S_{0}+g\stackrel{(1)}{S}_{0}+g^{2}\stackrel{%
(2)}{S}_{0}+\cdots ,  \label{19}
\end{equation}
and modify the gauge transformations (\ref{6}) (to be denoted by $\tilde{%
\delta}_{\epsilon }z^{A}$, $\tilde{\delta}_{\epsilon }u^{a_{0}}$) in such a
way that the deformed gauge transformations leave invariant the new action 
\begin{equation}
\frac{\delta \tilde{S}_{0}}{\delta z^{A}}\tilde{\delta}_{\epsilon }z^{A}+%
\frac{\delta \tilde{S}_{0}}{\delta u^{a_{0}}}\tilde{\delta}_{\epsilon
}u^{a_{0}}=0.  \label{20}
\end{equation}
Consequently, the deformation of the action (\ref{5}) and of the gauge
transformations (\ref{6}) produces a deformation of the first-class
constraints, first-class Hamiltonian and structure functions like 
\begin{equation}
G_{a_{0}}\rightarrow \gamma _{a_{0}}=G_{a_{0}}+g\stackrel{(1)}{\gamma }%
_{a_{0}}+g^{2}\stackrel{(2)}{\gamma }_{a_{0}}+\cdots ,  \label{14}
\end{equation}
\begin{equation}
H_{0}\rightarrow H=H_{0}+g\stackrel{(1)}{H}+g^{2}\stackrel{(2)}{H}+\cdots ,
\label{15}
\end{equation}
\begin{equation}
V_{\;\;b_{0}}^{a_{0}}\rightarrow \tilde{V}_{\;\;b_{0}}^{a_{0}}=V_{\;%
\;b_{0}}^{a_{0}}+g\stackrel{(1)}{V}_{\;\;b_{0}}^{a_{0}}+g^{2}\stackrel{(2)}{V%
}_{\;\;b_{0}}^{a_{0}}+\cdots ,  \label{16}
\end{equation}
\begin{equation}
C_{\;\;b_{0}c_{0}}^{a_{0}}\rightarrow \tilde{C}_{\;\;b_{0}c_{0}}^{a_{0}}=C_{%
\;\;b_{0}c_{0}}^{a_{0}}+g\stackrel{(1)}{C}_{\;\;b_{0}c_{0}}^{a_{0}}+g^{2}%
\stackrel{(2)}{C}_{\;\;b_{0}c_{0}}^{a_{0}}+\cdots ,  \label{17}
\end{equation}
such that the deformed gauge algebra becomes 
\begin{equation}
\left[ \gamma _{a_{0}},\gamma _{b_{0}}\right] =\gamma _{c_{0}}\tilde{C}%
_{\;\;a_{0}b_{0}}^{c_{0}},\;\left[ H,\gamma _{a_{0}}\right] =\gamma _{b_{0}}%
\tilde{V}_{\;\;a_{0}}^{b_{0}}.  \label{18}
\end{equation}
In the meantime, we deform the reducibility relations, but we do not
explicitly write down these relations.

As the BRST charge and BRST-invariant Hamiltonian contain all the
information on the gauge structure of a given gauge theory, we can
reformulate the problem of introducing consistent interactions within the
Hamiltonian BRST context in terms of these two essential compounds. Indeed,
if the interaction can be consistently constructed, then the BRST charge of
the undeformed theory, $\stackrel{(0)}{\Omega }$, can be deformed such as to
be the BRST charge of the deformed theory, i.e., 
\begin{equation}
\stackrel{(0)}{\Omega }\rightarrow \Omega =\stackrel{(0)}{\Omega }+g%
\stackrel{(1)}{\Omega }+g^{2}\stackrel{(2)}{\Omega }+\cdots ,  \label{21}
\end{equation}
\begin{equation}
\left[ \Omega ,\Omega \right] =0.  \label{22}
\end{equation}
At the same time, the deformation of the BRST charge induces the deformation
of the BRST-invariant Hamiltonian of the undeformed theory, $\stackrel{(0)}{H%
}_{B}$, 
\begin{equation}
\stackrel{(0)}{H}_{B}\rightarrow H_{B}=\stackrel{(0)}{H}_{B}+g\stackrel{(1)}{%
H}_{B}+g^{2}\stackrel{(2)}{H}_{B}+\cdots ,  \label{23}
\end{equation}
in such a way that $H_{B}$ is the BRST-invariant Hamiltonian of the
interacting theory, i.e. 
\begin{equation}
\left[ H_{B},\Omega \right] =0.  \label{24}
\end{equation}
The equations (\ref{22}) and (\ref{24}) split accordingly the deformation
parameter as 
\begin{equation}
\left[ \stackrel{(0)}{\Omega },\stackrel{(0)}{\Omega }\right] =0,\;\left[ 
\stackrel{(0)}{H}_{B},\stackrel{(0)}{\Omega }\right] =0,  \label{25}
\end{equation}
\begin{equation}
2\left[ \stackrel{(0)}{\Omega },\stackrel{(1)}{\Omega }\right] =0,\;\left[ 
\stackrel{(0)}{H}_{B},\stackrel{(1)}{\Omega }\right] +\left[ \stackrel{(1)}{H%
}_{B},\stackrel{(0)}{\Omega }\right] =0,  \label{26}
\end{equation}
\begin{equation}
2\left[ \stackrel{(0)}{\Omega },\stackrel{(2)}{\Omega }\right] +\left[ 
\stackrel{(1)}{\Omega },\stackrel{(1)}{\Omega }\right] =0,\;\left[ \stackrel{%
(0)}{H}_{B},\stackrel{(2)}{\Omega }\right] +\left[ \stackrel{(1)}{H}_{B},%
\stackrel{(1)}{\Omega }\right] +\left[ \stackrel{(2)}{H}_{B},\stackrel{(0)}{%
\Omega }\right] =0,  \label{27}
\end{equation}
\[
\vdots 
\]
Equations (\ref{25}--\ref{27}) stand for the general equations of our
deformation procedure. The equations (\ref{25}) are checked by hypothesis.
Then, it appears the natural question whether the next equations possess or
not solution. This will be investigated in the next section.

\section{Solution to the general equations}

In order to prove that the equations (\ref{26}--\ref{27}), etc. possess
solution, we use the link between the antifield and Hamiltonian BRST
formalisms for constrained Hamiltonian systems \cite{18}. First-class
constrained Hamiltonian systems can be approached from the point of view of
the BRST formalism in two different manners. One is based on the
antibracket-antifield formulation \cite{6}--\cite{10}, while the other
relies on the standard Hamiltonian BRST treatment \cite{10}, \cite{13}--\cite
{17}. The starting point of the antibracket-antifield formalism is
represented by the invariance of the action (\ref{5}) under the gauge
transformations (\ref{6}). In agreement with the general prescriptions of
the antibracket-antifield procedure, we introduce the ghosts $\left( \eta
^{a_{k-1}},u^{a_{k}}\right) $, with $k=1,\ldots ,L$ and 
\begin{equation}
\varepsilon \left( \eta ^{a_{k}}\right) =\varepsilon
_{a_{k}}+k+1\;mod\;2,\;gh\left( \eta ^{a_{k}}\right) =k+1,\;k=0,\ldots L,
\label{7}
\end{equation}
\begin{equation}
\varepsilon \left( u^{a_{k}}\right) =\varepsilon
_{a_{k}}+k\;mod\;2,\;gh\left( u^{a_{k}}\right) =k,\;k=1,\ldots L,  \label{8}
\end{equation}
where $gh$ denotes the ghost number. The antifields associated with the
fields $\left( z^{A},u^{a_{0}},\eta ^{a_{k-1}},u^{a_{k}}\right) $ are
denoted by $\left( z_{A}^{*},u_{a_{0}}^{*},\eta
_{a_{k-1}}^{*},u_{a_{k}}^{*}\right) $ and display the properties $%
\varepsilon \left( antifield\right) =\varepsilon \left( field\right) +1$, $%
gh\left( antifield\right) =-gh\left( field\right) -1$. Up to terms that are
quadratic in the antifields, the solution to the master equation reads as 
\begin{eqnarray}
&&\stackrel{(0)}{S}=\int\limits_{t_{1}}^{t_{2}}dt\left( a_{A}\left( z\right) 
\dot{z}^{A}+\sum\limits_{k=0}^{L}u_{a_{k}}^{*}\dot{\eta}%
^{a_{k}}-H_{0}-G_{a_{0}}u^{a_{0}}+z_{A}^{*}\left[ z^{A},G_{a_{0}}\right]
\eta ^{a_{0}}-\right.  \nonumber  \label{9} \\
&&u_{a_{0}}^{*}V_{\;\;b_{0}}^{a_{0}}\eta ^{b_{0}}+\left( -\right)
^{\varepsilon _{b_{0}}+1}u_{a_{0}}^{*}C_{\;\;b_{0}c_{0}}^{a_{0}}\eta
^{c_{0}}u^{b_{0}}+\frac{1}{2}\left( -\right) ^{\varepsilon _{b_{0}}}\eta
_{a_{0}}^{*}C_{\;\;b_{0}c_{0}}^{a_{0}}\eta ^{c_{0}}\eta ^{b_{0}}+  \nonumber
\\
&&\left. \sum\limits_{k=0}^{L-1}\eta _{a_{k}}^{*}Z_{\;\;a_{k+1}}^{a_{k}}\eta
^{a_{k+1}}-\sum\limits_{k=1}^{L-1}u_{a_{k-1}}^{*}Z_{\;%
\;a_{k}}^{a_{k-1}}u^{a_{k}}+\ldots \right) .
\end{eqnarray}
The Hamiltonian point of view is based on extending the phase-space through
introducing the canonical pairs ghost-antighost $\left( \eta ^{a_{k}},{\cal P%
}_{a_{k}}\right) $, with $\left[ \eta ^{a_{k}},{\cal P}_{a_{k}}\right]
=\delta _{\;\;b_{k}}^{a_{k}}$ and $\varepsilon \left( {\cal P}%
_{a_{k}}\right) =\varepsilon _{a_{k}}+k+1$, $gh\left( {\cal P}%
_{a_{k}}\right) =k+1$. The BRST charge starts like 
\begin{equation}
\stackrel{(0)}{\Omega }=G_{a_{0}}\eta ^{a_{0}}+\frac{1}{2}\left( -\right)
^{\varepsilon _{b_{0}}}{\cal P}_{a_{0}}C_{\;\;b_{0}c_{0}}^{a_{0}}\eta
^{c_{0}}\eta ^{b_{0}}+\sum\limits_{k=0}^{L-1}{\cal P}_{a_{k}}Z_{\;%
\;a_{k+1}}^{a_{k}}\eta ^{a_{k+1}}+\cdots ,  \label{10}
\end{equation}
such that $\left[ \stackrel{(0)}{\Omega },\stackrel{(0)}{\Omega }\right] =0$%
. The BRST-invariant extension of $H_{0}$ 
\begin{equation}
\stackrel{(0)}{H}_{B}=H_{0}+{\cal P}_{a_{0}}V_{\;\;b_{0}}^{a_{0}}\eta
^{b_{0}}+\cdots ,  \label{11}
\end{equation}
satisfies the equation $\left[ \stackrel{(0)}{H}_{B},\stackrel{(0)}{\Omega }%
\right] =0$. By employing the identifications 
\begin{equation}
u_{a_{k}}^{*}={\cal P}_{a_{k}},\;k=0,\ldots ,L,  \label{12}
\end{equation}
and extending the Dirac bracket such that $\left[ \eta
^{a_{k}},u_{a_{k}}^{*}\right] =\delta _{\;\;b_{k}}^{a_{k}}$, we get that 
\begin{eqnarray}
&&\frac{1}{2}\left( \stackrel{(0)}{S},\stackrel{(0)}{S}\right)
=\int\limits_{t_{1}}^{t_{2}}dt\left( -\frac{d}{dt}\stackrel{(0)}{\Omega }%
-\left[ \stackrel{(0)}{H}_{B},\stackrel{(0)}{\Omega }\right] +\frac{1}{2}%
z_{A}^{*}\left[ z^{A},\left[ \stackrel{(0)}{\Omega },\stackrel{(0)}{\Omega }%
\right] \right] +\right.  \nonumber  \label{13} \\
&&\left. \frac{1}{2}\sum\limits_{k=0}^{L}\eta _{a_{k}}^{*}\left[ \eta
^{a_{k}},\left[ \stackrel{(0)}{\Omega },\stackrel{(0)}{\Omega }\right]
\right] +\frac{1}{2}\sum\limits_{k=0}^{L}\left[ \left[ \stackrel{(0)}{\Omega 
},\stackrel{(0)}{\Omega }\right] ,u_{a_{k}}^{*}\right] u^{a_{k}}\right) .
\end{eqnarray}
The deformations (\ref{21}) and (\ref{23}) induce a deformation of the
solution to the master equation 
\begin{equation}
\stackrel{(0)}{S}\rightarrow S=\stackrel{(0)}{S}+g\stackrel{(1)}{S}+g^{2}%
\stackrel{(2)}{S}+\cdots ,  \label{28}
\end{equation}
such that the equation (\ref{13}) for the deformed theory becomes 
\begin{eqnarray}
&&\frac{1}{2}\left( S,S\right) =\int\limits_{t_{1}}^{t_{2}}dt\left( -\frac{d%
}{dt}\Omega -\left[ H_{B},\Omega \right] +\frac{1}{2}z_{A}^{*}\left[
z^{A},\left[ \Omega ,\Omega \right] \right] +\right.  \nonumber  \label{29}
\\
&&\left. \frac{1}{2}\sum\limits_{k=0}^{L}\eta _{a_{k}}^{*}\left[ \eta
^{a_{k}},\left[ \Omega ,\Omega \right] \right] +\frac{1}{2}%
\sum\limits_{k=0}^{L}\left[ \left[ \Omega ,\Omega \right]
,u_{a_{k}}^{*}\right] u^{a_{k}}\right) .
\end{eqnarray}
The equation (\ref{29}) splits accordingly the deformation parameter as (\ref
{13}) and 
\begin{eqnarray}
&&\left( \stackrel{(0)}{S},\stackrel{(1)}{S}\right)
=\int\limits_{t_{1}}^{t_{2}}dt\left( -\frac{d}{dt}\stackrel{(1)}{\Omega }%
-\left[ \stackrel{(0)}{H}_{B},\stackrel{(1)}{\Omega }\right] -\left[ 
\stackrel{(1)}{H}_{B},\stackrel{(0)}{\Omega }\right] +z_{A}^{*}\left[
z^{A},\left[ \stackrel{(0)}{\Omega },\stackrel{(1)}{\Omega }\right] \right]
+\right.  \nonumber  \label{30} \\
&&\left. \sum\limits_{k=0}^{L}\eta _{a_{k}}^{*}\left[ \eta ^{a_{k}},\left[ 
\stackrel{(0)}{\Omega },\stackrel{(1)}{\Omega }\right] \right]
+\sum\limits_{k=0}^{L}\left[ \left[ \stackrel{(0)}{\Omega },\stackrel{(1)}{%
\Omega }\right] ,u_{a_{k}}^{*}\right] u^{a_{k}}\right) ,
\end{eqnarray}
\begin{eqnarray}
&&\left( \stackrel{(0)}{S},\stackrel{(2)}{S}\right) +\frac{1}{2}\left( 
\stackrel{(1)}{S},\stackrel{(1)}{S}\right)
=\int\limits_{t_{1}}^{t_{2}}dt\left( -\frac{d}{dt}\stackrel{(2)}{\Omega }%
-\left[ \stackrel{(0)}{H}_{B},\stackrel{(2)}{\Omega }\right] -\left[ 
\stackrel{(1)}{H}_{B},\stackrel{(1)}{\Omega }\right] -\right.  \nonumber
\label{31} \\
&&\left[ \stackrel{(2)}{H}_{B},\stackrel{(0)}{\Omega }\right]
+z_{A}^{*}\left[ z^{A},\left[ \stackrel{(0)}{\Omega },\stackrel{(2)}{\Omega }%
\right] +\frac{1}{2}\left[ \stackrel{(1)}{\Omega },\stackrel{(1)}{\Omega }%
\right] \right] +  \nonumber \\
&&\sum\limits_{k=0}^{L}\eta _{a_{k}}^{*}\left[ \eta ^{a_{k}},\left[ 
\stackrel{(0)}{\Omega },\stackrel{(2)}{\Omega }\right] +\frac{1}{2}\left[ 
\stackrel{(1)}{\Omega },\stackrel{(1)}{\Omega }\right] \right] +  \nonumber
\\
&&\left. \sum\limits_{k=0}^{L}\left[ \left[ \stackrel{(0)}{\Omega },%
\stackrel{(2)}{\Omega }\right] +\frac{1}{2}\left[ \stackrel{(1)}{\Omega },%
\stackrel{(1)}{\Omega }\right] ,u_{a_{k}}^{*}\right] u^{a_{k}}\right) ,
\end{eqnarray}
\[
\vdots 
\]
The last equations emphasize that the existence of $\stackrel{(1)}{S}$
guarantees the existence of $\stackrel{(1)}{\Omega }$ and $\stackrel{(1)}{H}%
_{B}$, the existence of $\stackrel{(2)}{S}$ guarantees the existence of $%
\stackrel{(2)}{\Omega }$ and $\stackrel{(2)}{H}_{B}$, and so on. Moreover,
the equations (\ref{26}--\ref{27}), etc. are equivalent to the equations $%
\left( \stackrel{(0)}{S},\stackrel{(1)}{S}\right) =0$, $\left( \stackrel{(0)%
}{S},\stackrel{(2)}{S}\right) +\frac{1}{2}\left( \stackrel{(1)}{S},\stackrel{%
(1)}{S}\right) =0$, etc. modulo imposing some appropriate boundary
conditions for $\Omega $ \cite{17}. On the other hand, the last equations
possess solution. The existence of such solutions was proved in \cite{5} on
behalf of the triviality of the antibracket in the cohomology. Thus, the
existence of the solutions in the antibracket proves the existence of the
solutions to (\ref{26}--\ref{27}), etc. In conclusion, we can construct
consistent interactions by means of the equations (\ref{26}--\ref{27}), etc.

In practical applications, as commonly required, the deformation should be
local, i.e., $\stackrel{(1)}{\Omega }$, $\stackrel{(2)}{\Omega }$, $%
\stackrel{(1)}{H}_{B}$, $\stackrel{(2)}{H}_{B}$, etc. should be local
functionals. Let $F_{1}=\int d^{D-1}xf_{1}$ and $F_{2}=\int d^{D-1}xf_{2}$
be two local functionals. Then, $\left[ F_{1},F_{2}\right] $ is local,
namely, there exists a local $\left[ f_{1},f_{2}\right] $ (but defined up to
a $\left( D-1\right) $-dimensional divergence), such that $\left[
F_{1},F_{2}\right] =\int d^{D-1}x\left[ f_{1},f_{2}\right] $. Thus, the
equations (\ref{26}--\ref{27}), etc. can be written as 
\begin{equation}
2\stackrel{(0)}{s}\stackrel{(1)}{\omega }=\partial ^{k}\stackrel{(1)}{j}%
_{k},\;\stackrel{(0)}{s}\stackrel{(1)}{h}_{B}+\left[ \stackrel{(0)}{h}_{B},%
\stackrel{(1)}{\omega }\right] =\partial ^{k}\stackrel{(1)}{m}_{k},
\label{32}
\end{equation}
\begin{equation}
2\stackrel{(0)}{s}\stackrel{(2)}{\omega }+\left[ \stackrel{(1)}{\omega },%
\stackrel{(1)}{\omega }\right] =\partial ^{k}\stackrel{(2)}{j}_{k},\;%
\stackrel{(0)}{s}\stackrel{(2)}{h}_{B}+\left[ \stackrel{(1)}{h}_{B},%
\stackrel{(1)}{\omega }\right] +\left[ \stackrel{(0)}{h}_{B},\stackrel{(2)}{%
\omega }\right] =\partial ^{k}\stackrel{(2)}{m}_{k},  \label{33}
\end{equation}
\[
\vdots 
\]
in terms of the integrands $\stackrel{(k)}{h}_{B}$ and $\stackrel{(k)}{%
\omega }$. In the above, $\stackrel{(0)}{s}$ stands for the undeformed BRST
symmetry. The formalism developed so far does not guarantee locality. For
instance, even if $\left[ \stackrel{(1)}{\Omega },\stackrel{(1)}{\Omega }%
\right] $ is $\stackrel{(0)}{s}$-exact, it is not granted that it is the
BRST variation of a local functional. Such locality problems appear also in
the Lagrangian deformation procedure \cite{5}. However, in the case of most
important applications \cite{5}, \cite{11}--\cite{12a}, the Lagrangian BRST
deformation procedure leads to local interactions. Thus, we expect that the
Hamiltonian BRST deformation treatment also outputs local vertices in
practical applications.

\section{Example}

Let us exemplify the prior procedure in the case of abelian Chern-Simons
model in three dimensions. We start with the Lagrangian action 
\begin{equation}
S_{0}\left[ A_{\mu }^{a}\right] =\frac{1}{2}\int d^{3}x\varepsilon ^{\mu \nu
\rho }k_{ab}A_{\mu }^{a}F_{\nu \rho }^{b},  \label{34}
\end{equation}
where $k_{ab}$ is a non-degenerate symmetric and constant matrix, while $%
F_{\nu \rho }^{b}=\partial _{\nu }A_{\rho }^{b}-\partial _{\rho }A_{\nu
}^{b}\equiv \partial _{\left[ \nu \right. }A_{\left. \rho \right] }^{b}$.
Performing the canonical analysis and eliminating the second-class
constraints (the independent variables are $A_{0}^{a}$, $\pi _{a}^{0}$ and $%
A_{k}^{a}$), we infer the first-class constraints $G_{1a}\equiv \pi
_{a}^{0}\approx 0$, $G_{2a}\equiv -\frac{1}{2}\varepsilon
^{0ik}k_{ab}F_{ik}^{b}\approx 0$ and the first-class Hamiltonian $%
H_{0}=-2\int d^{2}xA_{0}^{a}G_{2a}$. The non-vanishing fundamental Dirac
brackets read as $\left[ A_{0}^{a},\pi _{b}^{0}\right] =\delta _{\;\;b}^{a}$%
, $\left[ A_{k}^{a},A_{j}^{b}\right] =\frac{1}{2}\varepsilon _{0kj}k^{ab}$,
hence the BRST charge takes the simple form 
\begin{equation}
\stackrel{(0)}{\Omega }=\int d^{2}x\left( \pi _{a}^{0}\eta _{1}^{a}-\frac{1}{%
2}\varepsilon ^{0ik}k_{ab}F_{ik}^{b}\eta _{2}^{a}\right) ,  \label{38}
\end{equation}
where $k^{ab}$ is the inverse of $k_{ab}$, and $\left( \eta _{1}^{a},\eta
_{2}^{a}\right) $ stand for the fermionic ghost number one ghosts. Thus, the
BRST operator $\stackrel{(0)}{s}$ splits as $\stackrel{(0)}{s}=\delta
+\gamma $, where $\delta $ is the Koszul-Tate differential and $\gamma $
represents the longitudinal exterior derivative along the gauge orbits.
Then, we have 
\begin{equation}
\delta A_{0}^{a}=0,\;\delta \pi _{a}^{0}=0,\;\delta A_{k}^{a}=0,\;\delta
\eta _{1}^{a}=\delta \eta _{2}^{a}=0,  \label{39}
\end{equation}
\begin{equation}
\delta {\cal P}_{1a}=-\pi _{a}^{0},\;\delta {\cal P}_{2a}=\frac{1}{2}%
\varepsilon ^{0ik}k_{ab}F_{ik}^{b},  \label{40}
\end{equation}
\begin{equation}
\gamma A_{0}^{a}=\eta _{1}^{a},\;\gamma \pi _{a}^{0}=0,\;\gamma A_{k}^{a}=%
\frac{1}{2}\partial _{k}\eta _{2}^{a},\;\gamma \eta _{1}^{a}=\gamma \eta
_{2}^{a}=0,  \label{41}
\end{equation}
\begin{equation}
\gamma {\cal P}_{1a}=\gamma {\cal P}_{2a}=0.  \label{42}
\end{equation}
Now, we solve the former equation from (\ref{32}). In view of this, we
develop $\stackrel{(1)}{\omega }$ accordingly the antighost number 
\begin{equation}
\stackrel{(1)}{\omega }=\stackrel{(1)}{\omega }_{0}+\stackrel{(1)}{\omega }%
_{1}+\cdots +\stackrel{(1)}{\omega }_{\Delta },\;antigh\left( \stackrel{(1)}{%
\omega }_{\Delta }\right) =\Delta ,\;gh\left( \stackrel{(1)}{\omega }%
_{\Delta }\right) =1,  \label{43}
\end{equation}
where the last term in (\ref{43}) can be assumed to be annihilated by $%
\gamma $. As $pgh\left( \stackrel{(1)}{\omega }_{\Delta }\right) =\Delta +1$%
, we can represent $\stackrel{(1)}{\omega }_{\Delta }$ under the form 
\begin{equation}
\stackrel{(1)}{\omega }_{\Delta }=\mu _{a_{1}\cdots a_{\Delta +1}}\eta
_{2}^{a_{1}}\cdots \eta _{2}^{a_{\Delta +1}}.  \label{44}
\end{equation}
With this choice, it results that the $\gamma $-invariant coefficient $\mu
_{a_{1}\cdots a_{\Delta +1}}$ belongs to $H_{\Delta }\left( \delta |\tilde{d}%
\right) $, i.e., is solution to the equation 
\begin{equation}
\delta \mu _{a_{1}\cdots a_{\Delta +1}}+\partial _{k}b_{a_{1}\cdots
a_{\Delta +1}}^{k}=0,  \label{45}
\end{equation}
for some $b_{a_{1}\cdots a_{\Delta +1}}^{k}$, where $\tilde{d}%
=dx^{i}\partial _{i}$. Using the result from \cite{25} adapted to the
Hamiltonian context, it follows that $H_{\Delta }\left( \delta |\tilde{d}%
\right) $ vanish for $\Delta \geq 2$ , hence $\stackrel{(1)}{\omega }=%
\stackrel{(1)}{\omega }_{0}+\stackrel{(1)}{\omega }_{1}$, with $\stackrel{(1)%
}{\omega }_{1}=\frac{1}{2}\mu _{ab}\eta _{2}^{a}\eta _{2}^{b}$, where $\mu
_{ab}$ from $H_{1}\left( \delta |\tilde{d}\right) $. A general
representative of $H_{1}\left( \delta |\tilde{d}\right) $ is of the type $%
\mu _{ab}=C_{\;\;ab}^{c}{\cal P}_{2c}$, where $C_{\;\;ab}^{c}$ are some
constants, antisymmetric in the lower indices, $C_{\;\;ab}^{c}=-C_{\;%
\;ba}^{c}$. The necessity for $C_{\;\;ab}^{c}$ to be constant results from
the equation that must be satisfied by $\mu _{ab}$, namely, $\delta \mu
_{ab}=\partial _{k}\left( C_{\;\;ab}^{c}\varepsilon
^{0kj}k_{cd}A_{j}^{d}\right) $. In this way, we obtained that $\stackrel{(1)%
}{\omega }_{1}=\frac{1}{2}C_{\;\;ab}^{c}{\cal P}_{2c}\eta _{2}^{a}\eta
_{2}^{b}$. The former equation from (\ref{32}) at antighost number zero
reads as $\delta \stackrel{(1)}{\omega }_{1}+\gamma \stackrel{(1)}{\omega }%
_{0}=\partial _{k}m^{k}$, which further yields $\stackrel{(1)}{\omega }%
_{0}=C_{\;\;ad}^{c}k_{cb}\varepsilon ^{0kj}A_{k}^{a}A_{j}^{d}\eta _{2}^{b}$.
In this manner, we inferred $\stackrel{(1)}{\omega }=C_{\;\;ab}^{c}\left( 
\frac{1}{2}{\cal P}_{2c}\eta _{2}^{a}\eta _{2}^{b}+k_{cd}\varepsilon
^{0kj}A_{k}^{a}A_{j}^{b}\eta _{2}^{d}\right) $. Simple computation leads to 
\begin{eqnarray}
&&\left[ \stackrel{(1)}{\Omega },\stackrel{(1)}{\Omega }\right] =\int
d^{2}x\left( -\frac{1}{3}C_{\;\;\left[ nc\right. }^{m}C_{\;\;\left.
ab\right] }^{c}{\cal P}_{2m}\eta _{2}^{a}\eta _{2}^{b}\eta _{2}^{n}-\right. 
\nonumber  \label{49} \\
&&\left. \varepsilon ^{0ij}k_{ad}C_{\;\;\left[ bc\right. }^{d}C_{\;\;\left.
ne\right] }^{c}\eta _{2}^{a}\eta _{2}^{b}A_{i}^{n}A_{j}^{e}\right) .
\end{eqnarray}
The last relation shows that $\left[ \stackrel{(1)}{\Omega },\stackrel{(1)}{%
\Omega }\right] $ cannot be written like a $\stackrel{(0)}{s}$-exact modulo $%
\tilde{d}$ local functional. For this reason it is necessary to have $\left[ 
\stackrel{(1)}{\Omega },\stackrel{(1)}{\Omega }\right] =0$. This condition
takes place if and only if $C_{\;\;\left[ nc\right. }^{m}C_{\;\;\left.
ab\right] }^{c}=0$, so if and only if the constants verify the Jacobi
identity. This further implies $\stackrel{(k)}{\Omega }=0$,\ $k\geq 2$. The
deformed BRST charge takes the final form 
\begin{eqnarray}
&&\Omega =\int d^{2}x\left( \pi _{a}^{0}\eta _{1}^{a}-\varepsilon
^{0ik}k_{ca}\left( \frac{1}{2}F_{ik}^{c}-gC_{\;\;bd}^{c}A_{i}^{b}A_{k}^{d}%
\right) \eta _{2}^{a}+\right.  \nonumber  \label{50} \\
&&\left. \frac{1}{2}gC_{\;\;ab}^{c}{\cal P}_{2c}\eta _{2}^{a}\eta
_{2}^{b}\right) ,
\end{eqnarray}
so it is clearly a local functional.

Now, we derive the deformed BRST-invariant Hamiltonian. The BRST-invariant
Hamiltonian for the free theory is given by $\stackrel{(0)}{H}%
_{B}=H_{0}+2\int d^{2}x\eta _{1}^{a}{\cal P}_{2a}$. Consequently, we find 
\begin{equation}
\left[ \stackrel{(0)}{h}_{B},\stackrel{(1)}{\omega }\right]
=-2C_{\;\;ab}^{c}k_{cd}\varepsilon ^{0ij}A_{j}^{b}\left( \eta
_{1}^{d}A_{i}^{a}+\eta _{2}^{d}\partial _{i}A_{0}^{a}\right) -2C_{\;\;ab}^{c}%
{\cal P}_{2c}\eta _{2}^{a}\eta _{1}^{b}.  \label{52}
\end{equation}
Then, the solution of the latter equation in (\ref{32}) reads as 
\begin{equation}
\stackrel{(1)}{h}_{B}=2C_{\;\;ab}^{c}\left( k_{cd}\varepsilon
^{0ij}A_{0}^{d}A_{i}^{a}A_{j}^{b}+A_{0}^{b}{\cal P}_{2c}\eta _{2}^{a}\right)
.  \label{52a}
\end{equation}
Straightforward computation leads to $\left[ \stackrel{(1)}{H}_{B},\stackrel{%
(1)}{\Omega }\right] =0$, hence the latter equation from (\ref{33}) is
satisfied with $\stackrel{(2)}{h}_{B}=0$. Therefore, the higher-order
deformation equations for the BRST-invariant Hamiltonian are verified with $%
\stackrel{(3)}{H}_{B}=\stackrel{(4)}{H}_{B}=\cdots =0$. Thus, the complete
deformed BRST invariant Hamiltonian reads as 
\begin{equation}
H_{B}=2\int d^{2}x\left( -A_{0}^{a}\varepsilon ^{0ik}k_{ca}\left( \frac{1}{2}%
F_{ik}^{c}-gC_{\;\;bd}^{c}A_{i}^{b}A_{k}^{d}\right) +\left( \eta
_{1}^{a}-gC_{\;\;cb}^{a}A_{0}^{b}\eta _{2}^{c}\right) {\cal P}_{2a}\right) ,
\label{53}
\end{equation}
and is a local functional, too. With the help of (\ref{50}) and (\ref{53})
we identify the new gauge theory. From the antighost-independent terms in (%
\ref{50}) we observe that the deformation of the BRST charge implies the
deformed first-class constraints 
\begin{equation}
\gamma _{2a}\equiv -\varepsilon ^{0ik}k_{ca}\left( \frac{1}{2}%
F_{ik}^{c}-gC_{\;\;bd}^{c}A_{i}^{b}A_{k}^{d}\right) \approx 0,  \label{53a}
\end{equation}
the remaining constraints being undeformed. The term $\frac{1}{2}%
gC_{\;\;ab}^{c}{\cal P}_{2c}\eta _{2}^{a}\eta _{2}^{b}$ shows that the new
constraint functions form a Lie algebra, i.e., 
\begin{equation}
\left[ \gamma _{2a},\gamma _{2b}\right] =C_{\;\;ab}^{c}\gamma _{2c}.
\label{53c}
\end{equation}
On the other hand, the antighost-independent piece in (\ref{53}) 
\begin{equation}
H=-2\int d^{2}xA_{0}^{a}\varepsilon ^{0ik}k_{ca}\left( \frac{1}{2}%
F_{ik}^{c}-gC_{\;\;bd}^{c}A_{i}^{b}A_{k}^{d}\right) ,  \label{53b}
\end{equation}
is precisely the first-class Hamiltonian of the deformed theory. The
components linear in the antighosts from (\ref{53}) indicate that the Dirac
brackets among the new first-class Hamiltonian and the new constraint
functions are modified as $\left[ H,\gamma _{2a}\right]
=-C_{\;\;ab}^{c}A_{0}^{b}\gamma _{2c}$. Thus, the resulting first-class
theory is nothing but the nonabelian version of the Chern-Simons model in
three dimensions, described by the local Lagrangian action 
\begin{equation}
\bar{S}_{0}\left[ A_{\mu }^{a}\right] =\int d^{3}x\varepsilon ^{\mu \nu \rho
}A_{\mu }^{a}\left( \frac{1}{2}k_{ab}F_{\nu \rho }^{b}-\frac{2}{3}%
gC_{abc}A_{\nu }^{b}A_{\rho }^{c}\right) ,  \label{54}
\end{equation}
where $C_{abc}=C_{\;\;\left[ bc\right. }^{d}k_{\left. a\right] d}$. As the
first-class constraints generate gauge transformations, from the
deformations (\ref{53a}) and (\ref{53c}) we can conclude that the added
interactions involved with (\ref{53b}) modify both the gauge transformations
and their gauge algebra.

\section{Conclusion}

To conclude with, in this letter we have presented a Hamiltonian BRST
approach to the construction of consistent interactions among fields with
gauge freedom. Our procedure reformulates the problem of constructing
Hamiltonian consistent interactions as a deformation problem of the BRST
charge and BRST-invariant Hamiltonian of a given ``free'' theory. We have
derived the general equations from the Hamiltonian BRST deformation method,
and proved that they possess solution. Next, we have written down the local
version of these equations and discussed on the locality of their solutions.
Finally, the general theory was exemplified in the case of the Chern-Simons
model in three dimensions. We think that our approach together with the
general results in \cite{12c} might be successfully applied to computing
local BRST cohomologies for those theories whose Lagrangian version is more
intricate than the Hamiltonian one.

\section*{Acknowledgment}

This work has been supported by a Romanian National Council for Academic
Scientific Research (CNCSIS) grant.

\end{document}